\documentclass{aa} 
\usepackage{natbib}
\usepackage{graphicx}
\usepackage{txfonts}
\begin{document}

\title{A Thorough Investigation of Distance and Age of the 
Pulsar Wind Nebula 3C\,58}

\titlerunning{Distance and Age of PWN 3C\,58}
\authorrunning{R. Kothes}

\author{R. Kothes \inst{1}}

\institute{National Research Council Herzberg,
              Dominion Radio Astrophysical Observatory,
              P.O. Box 248, Penticton, British Columbia, V2A 6J9, Canada}
   \date{Received ; accepted }
   \offprints{R. Kothes}


\abstract
{A growing number of researchers present evidence that the pulsar wind nebula 
3C\,58 is much older than predicted by its proposed connection to the historical 
supernova of A.D.~1181. There is also a great diversity of arguments. The
strongest of these arguments rely heavily on the 
assumed distance of 3.2\,kpc determined with H~{\sc{i}} absorption measurements.}
{This publication aims at determining a more accurate distance for 3C\,58 and
re-evaluating the arguments for a larger age.}
{I have re-visited the distance determination of 3C\,58 based on new H~{\sc{i}} data from 
the Canadian Galactic Plane Survey and our recent improvements in the knowledge of the
rotation curve of the outer Milky Way Galaxy. I have also used newly determined 
distances to objects in the neighbourhood, which are based on direct 
measurements by trigonometric parallax.}
{I have derived a new more reliable distance estimate of 2\,kpc for 3C\,58. This makes the 
connection between the pulsar wind nebula and the historical event from A.D.~1181 once 
again much more viable.}
{}

\keywords{ISM: individual (3C\,58), ISM: supernova remnants, pulsars: individual (J\,0205+6449)}

   \maketitle

\section{Introduction}

Of all the supernova remnants (SNRs) that are linked to historically observed 
supernova events the most disputed one is probably the connection between 
the pulsar wind nebula (PWN) 3C\,58 and the supernova explosion observed 
in A.D.~1181 by Chinese and Japanese astronomers. \citet{step71} and 
\citet{step02} claim that there is a 
high probability for a connection between the Guest Star from A.D.~1181 and the 
supernova explosion that created 3C\,58. The length of the visibility of the 
Guest Star and the similarity in the description in independent sources 
indicate a supernova explosion. The lack of any other supernova remnant 
candidate in the area strongly supports the identification of 3C\,58 with 
this historical supernova explosion. Although it is difficult to escape 
this argument, strong evidence against such a young age for 
3C\,58 has been mounting with a great diversity of support from expansion 
studies, theoretical
modeling of the evolutionary path, and comparisons of the pulsar's and the
pulsar wind nebula's characteristics to other SNRs and PWNe of known age
\citep[for a list of the main arguments see e.g.][Table 3]{fese08}.

Many of the strong arguments against a young age for 3C\,58 rely heavily 
on the assumed distance of 3.2~kpc \citep{robe93}. This distance was 
determined kinematically from H~{\sc{i}} absorption measurements by 
comparing the resulting systemic velocity for 3C\,58 with a flat 
rotation curve for the Outer Galaxy. For 
Perseus arm objects in particular this can lead to a significant overestimate of the 
distance. A spiral shock in the Perseus arm is ``pushing'' 
objects, including H~{\sc{ii}} regions, SNRs, and PWNe towards the Galactic centre, 
giving them -- from our point of view -- a 
higher negative radial velocity, which makes them appear to 
be farther away than they actually are \citep{robe72}. Examples of this 
effect on distance estimates of SNRs and PWNe can be found, e.g.
in \citet{koth02,koth03} and \citet{fost04} and most recently in \citet{koth12}. SNRs and PWNe
in particular rely heavily on kinematic distance determinations, since 
their distances cannot easily be determined
directly or by relating them to embedded stars as in the case for 
H~{\sc{ii}} regions.  

In this paper I discuss a new distance estimate for 3C\,58.
Based on this newly determined distance I will re-evaluate 
evidence presented in the literature against the historical connection between 
3C\,58 and the Guest Star of A.D. 1181. I will show that the new distance 
changes some of the PWN's characteristics quite dramatically. This leads
once more to a higher probability for its historical connection. I will
describe the H~{\sc{i}} observations of the Canadian Galactic Plane Survey in 
Section 2. H~{\sc{i}} emission around 3C\,58, its newly determined H~{\sc{i}} 
absorption characteristics, and the resulting systemic velocity are discussed 
in Section 3 and a thorough evaluation of the distance is presented in
Section 4. In section 5 I will re-visit all arguments contra and pro a young
age, develop a model in Section 6 for the evolution of 3C\,58 that 
explains all observations, and give a short summary in Section 7.
Preliminary results of this project were published in \citet{koth10}.

\section{Observations}

\begin{figure*}[!ht]
\begin{center}
  \includegraphics[bb = 100 260 475 625,width=0.80\textwidth,clip]{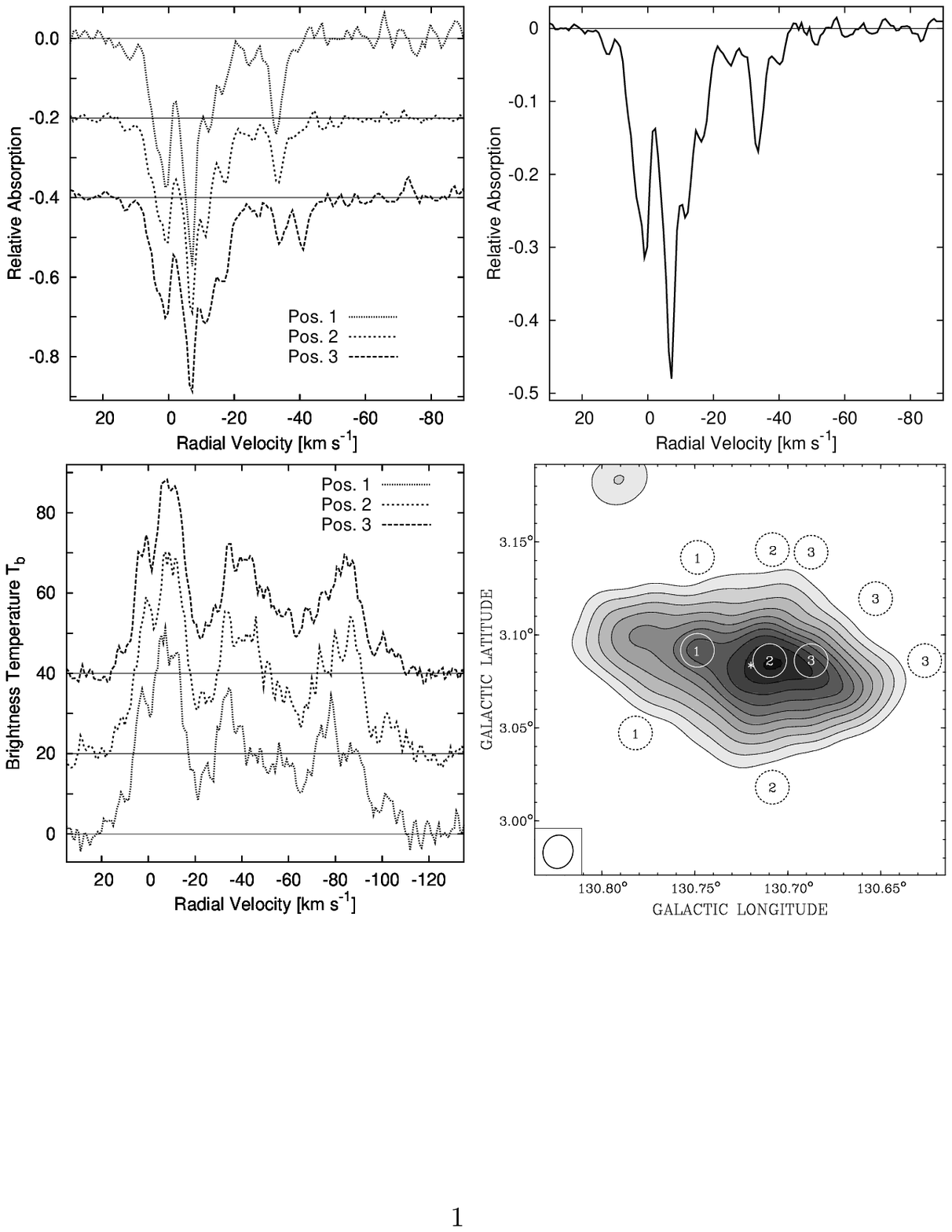}
\end{center}
 \caption{{\bf Top}: H~{\sc{i}} absorption profiles of 3C\,58. The relative absorption 
     of the radio continuum signal at 1420~MHz is displayed as a function of radial 
     velocity $v_{LSR}$. Left: Absorption profiles of 3 different positions are 
     displayed as indicated in the bottom right panel. Right: Profile
     calculated using all pixels on 3C\,58 with $T_b \ge 100$\,K. Each pixel was 
     weighted by its intensity.
     {\bf Bottom}: Left: H~{\sc{i}} emission profiles used to determine the 
     absorption profiles in the top left. These were averaged
     over the positions marked in the right panel. Right: Continuum image of
     3C\,58 taken from the CGPS. The "on" and "off" positions used to calculate the 
     H~{\sc{i}} absorption profiles are indicated by solid and dashed circles,
     respectively.}
   \label{abs}
\end{figure*}

The data for this new H~{\sc{i}} absorption study towards 3C\,58
were obtained with the Synthesis Telescope at the Dominion 
Radio Astrophysical Observatory \citep[DRAO,][]{land00} as part of the Canadian 
Galactic Plane Survey \citep[CGPS,][]{tayl03}. Single antenna data were 
incorporated into the interferometer maps after suitable filtering in the
Fourier domain. This assures the accurate representation of atomic hydrogen
emission from the largest 
structures down to the resolution limit of about 1\arcmin. The low spatial 
frequency data were drawn from the Low-Resolution DRAO Survey (LRDS) of the CGPS 
region, which was observed with the 26-m radio telescope at DRAO 
\citep{higg00}. The resolution in the final data sets varies slightly across the 
images as about $1\arcmin \times 1\arcmin {\rm cosec(Declination)}$. At the 
centre of 3C\,58 the resolution of the CGPS H~{\sc{i}} data is $58\arcsec \times 65\arcsec$ 
at an angle of $75\degr$ for the major axis (counter-clockwise from the 
Galactic longitude axis). The RMS noise is about 3\,K $T_B$ in each velocity 
channel of width 0.82446\,km~s$^{-1}$. The frequency band used for the CGPS 
H~{\sc{i}} observations is 1\,MHz wide, which gives a velocity coverage of about 
210\,km\,\,s$^{-1}$ at a velocity resolution of 1.3\,km~s$^{-1}$. In this area 
of the Milky Way Galaxy the CGPS H~{\sc{i}} data are centred at a radial 
velocity of $-60$\,km~s$^{-1}$ relative to the local standard of rest.

\section{Results}

\subsection{H~{\sc{i}} Absorption Towards 3C\,58}

The first attempt to derive an H~{\sc{i}} absorption distance to 3C\,58 was carried
out by \citet{will73}, who found absorption at negative velocities down to 
$-95$\,km\,\,s$^{-1}$, which
was translated to a distance of 8.2~kpc. However, those observations were obtained
with a single-dish telescope and the ``off position'' to determine the background 
H~{\sc{i}} emission for the absorption profile
was taken $9\arcmin$ away from the centre of 3C\,58. \citet{gree82} who
did the first absorption study towards 3C\,58 with an interferometer
argue that most of the absorption components at high negative velocities in the work
of \citet{will73} are artificial due to the highly structured interstellar medium. They
found no absorption beyond $-34$~km~s$^{-1}$ and determined a systemic velocity of
about $-37$~km~s$^{-1}$, which they translated with a Schmidt rotation model
\citep{schm65} for our Galaxy to a distance of 2.6~kpc. 
The latest distance
determination for 3C\,58 by H~{\sc{i}} absorption measurements resulted in a 
systemic velocity of
$\sim -38$\,km~s$^{-1}$ and a Perseus arm location \citep{robe93}. This was translated 
with a flat rotation model for our Galaxy
and the IAU supported values for the Sun's Galacto-centric distance of 
$R_\odot = 8.5$~kpc and the Sun's circular motion around the Galactic centre of
$v_\odot = 220$\,km~s$^{-1}$ to a distance of 3.2~kpc.

\begin{figure*}[!ht]
\begin{center}
 \includegraphics[bb = 35 125 555 480,width=0.85\textwidth,clip]{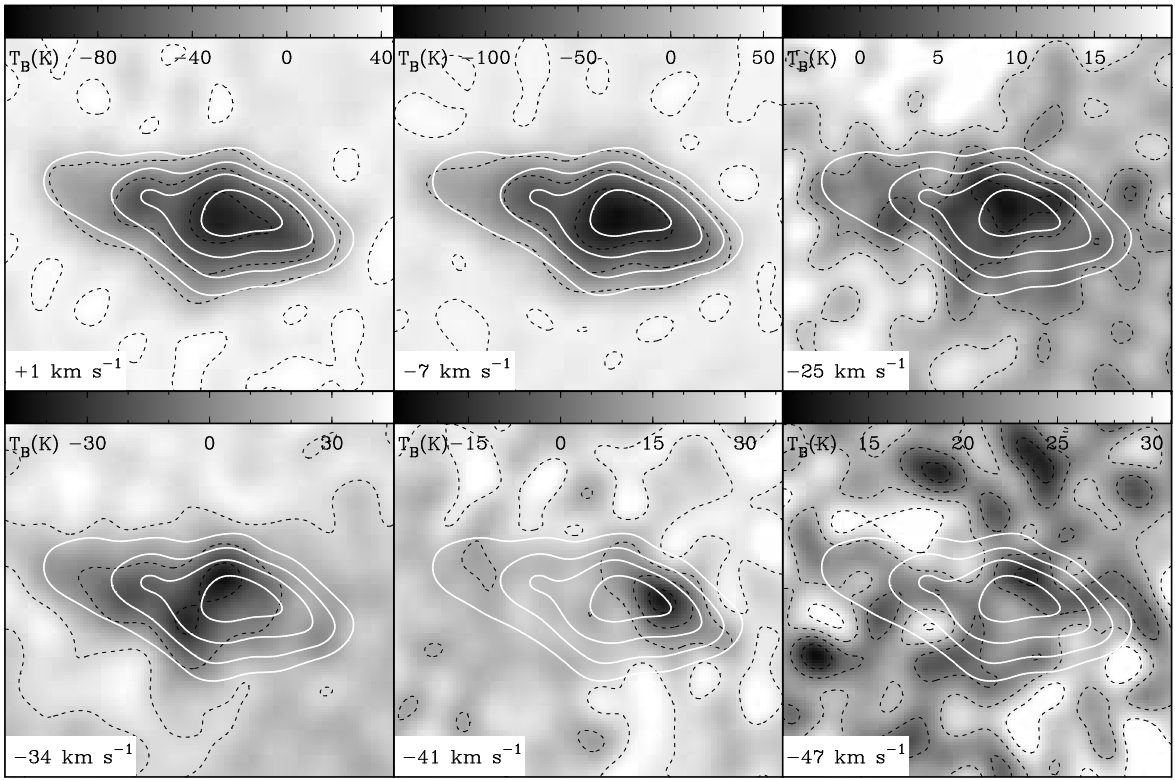}
\end{center}
\caption{H~{\sc{i}} channel maps of 3C\,58 taken at the peaks in the 
absorption spectra at +1 (top left), $-$7 (top centre), $-$34 (bottom left), and
$-$41~km\,s$^{-1}$ (bottom centre). The right panels show 10 channels averaged over
the inter-arm velocities between the Local and Perseus arm (top) and 10 channels
averaged over the velocities just outside the last absorption feature.
The continuum emission of 3C\,58 is indicated
by the white contours. The black dashed lines represent the H~{\sc{i}} contours at the 
levels indicated by the labels of the colour bars.}
\label{abs2}
\end{figure*}

I used the H~{\sc{i}} data of the CGPS to conduct a new H~{\sc{i}} absorption study 
towards 3C\,58 to investigate not only the integrated absorption spectrum, but 
also any possible significant changes over the PWN: if absorbing clouds are
small then the absorption by those clouds may be washed out in an integrated
profile. This could cause the loss of important information about the 
velocity range in which 3C\,58 is absorbed. In Fig.~\ref{abs} I show three absorption
profiles towards individual positions on 3C\,58. To calculate the displayed
spectra I subtracted the averaged emission profiles as indicated on the map
in the lower right panel of Fig.~\ref{abs}. For the integrated absorption
profile I averaged over all pixels on 3C\,58 that display a surface brightness in excess
of 100\,K~$T_b$ in the 1420\,MHz continuum maps of the CGPS. Each pixel was
weighted by its brightness above the background. For the background 
emission profile the average over all locations marked in Fig.~\ref{abs} was
calculated. The resulting profiles are shown in Fig.~\ref{abs}. 
H~{\sc{i}} channel maps
taken at the major peaks of the integrated absorption spectrum are displayed in 
Fig.~\ref{abs2}. These peaks can be found at $+1$ and $-7$\,km~s$^{-1}$ which
represents local gas, at $-34$\,km~s$^{-1}$ representing the Perseus spiral arm
and for a small area in the right part of 3C\,58 at $-41$\,km~s$^{-1}$.
In addition I added two maps averaged over 10 velocity
channels in the 
inter-arm area between the Local and the Perseus arm and 10 velocity channels
just beyond the last visible absorption component at $-41$\,km\,\,s$^{-1}$.

The integrated absorption spectrum is almost identical to the one published 
by \citet{robe93}. Every peak and wiggle is perfectly reproduced, which is
not surprising since 3C\,58 is a very bright source. At the local velocities and 
in the inter-arm area between the Local arm and Perseus arm the absorption looks quite
constant over the PWN. However, I found that the weak peak, seen as a faint wiggle 
at around $-41$~km~s$^{-1}$ is actually a real absorption feature. It is not 
seen over the entire PWN, but only on the right hand side at position 3 
(see Figs.~\ref{abs} and \ref{abs2}).
This is likely the reason why this feature was not obvious in previously
published integrated absorption spectra. The Perseus arm absorption component
around $-34$\,km\,\,s$^{-1}$ seems to be confined to the left hand side of the
PWN and is absent from the lower right. The reason for this is probably the structured 
interstellar medium. In the local medium clouds have a bigger angular size than
in the Perseus arm simply because they are much closer. Hence the absorbing clouds in
the Perseus arm show structure on smaller angular scale.

\begin{figure*}[!ht]
\begin{center}
 \includegraphics[bb = 25 25 535 740,width=0.85\textwidth,clip]{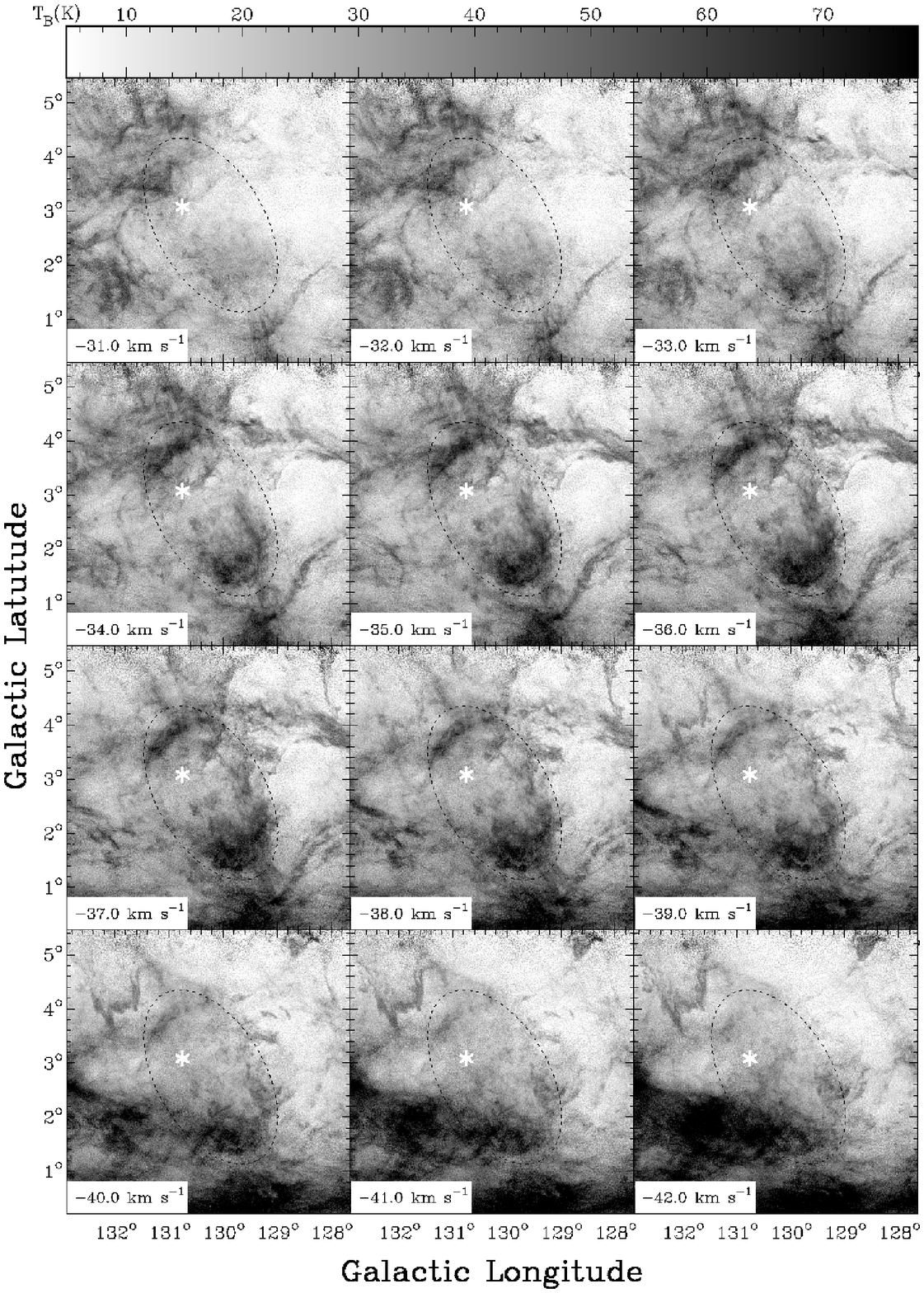}
\end{center}
\caption{H~{\sc{i}} channel maps of the area around 3C\,58. Each channel is 
1.0~km\,s$^{-1}$ wide and the centre velocity is indicated. The location 
of 3C\,58 is marked by the white asterisk. The black ellipse indicates the
approximate extend of the bipolar bubble discovered by \citet{wall94}.}
\label{hiemi}
\end{figure*}

\subsection{H~{\sc{i}} Emission Around 3C\,58}

\citet{wall94} discussed the possible association of 3C\,58 with a large
bi-polar interstellar bubble which they detected in low resolution H~{\sc{i}} 
data. 
An inspection of the CGPS H~{\sc{i}} data sets reveals that this double-shell 
structure is actually a relatively smooth object (see Fig.~\ref{hiemi}). 
This 
bubble is visible from about $-33$ to $\bf -39$\,km\,s$^{-1}$ and its 
structure is 
essentially unchanging over that entire velocity range. No caps are obvious in 
emission, neither at more nor at less negative velocities. The systemic velocity 
of this structure is about $-36$\,km\,s$^{-1}$. It seems to appear at 
about $-32$ to $-33$\,km\,s$^{-1}$ without any dynamical connection to all the
other filaments that are visible in this velocity range and disappears the same
way at about $-39$ to $-40$\,km\,s$^{-1}$.
The fact that this bubble seems to be stationary through many velocity channels 
and does not show any cap moving towards or away from us does not contradict 
its interpretation as an expanding bubble. \citet{cazz05} simulated H~{\sc{i}} 
observations of expanding stellar wind bubbles and found that this is 
not unexpected. 
Stationary rings or shells and weak or undetectable caps are the result of velocity 
dispersion. A very nice example for an actual observation of such a
structure is the stellar wind bubble around the SNR 3C434.1, where the receding
cap is seen absorbing warm background H~{\sc{i}} emission \citep{fost04}.

3C\,58 can be found projected onto the top left area of this bubble. If 3C\,58 
was located inside this structure, the last absorption feature at about 
$-41$\,km\,s$^{-1}$ must be produced by the cap of this bubble which is 
expanding towards us. After correction for the off-centre position of 3C\,58 this would result
in an expansion velocity of 6 to 7\,km~s$^{-1}$ for the bubble.

\subsection{The Systemic Velocity of 3C\,58}

\citet{robe93} found a systemic velocity of about 
$v_{LSR} = -38$\,km\,s$^{-1}$ (LSR = Local Standard of Rest), which they translated 
to a distance of 3.2\,kpc, in the Perseus arm.
This systemic velocity was confirmed with the discovery of the large bubble 
by \citet{wall94}. The CGPS H~{\sc{i}} absorption profile (Fig.~\ref{abs}) shows an 
additional absorption feature at about $-41$\,km\,s$^{-1}$ and a systemic 
velocity of about $-36$\,km~s$^{-1}$ for the bubble. There is no absorption
at velocities beyond $-41$\,km~s$^{-1}$. 
In Fig.~\ref{abs2} the map averaged
around $-25$\,km~s$^{-1}$, representative of the inter-arm between the Local arm and
the Perseus arm shows significant absorption. However, the map averaged over velocities 
beyond $-41$\,km~s$^{-1}$ is free of absorption even though this velocity range
should represent a location inside a spiral arm with denser gas than 
the inter-arm around $-25$\,km~s$^{-1}$. 
This confirms the Perseus arm location of 3C\,58. 3C\,58 is most likely
located inside the bubble shown in Fig.~\ref{hiemi} and the last absorption feature,
which is only partly visible in the absorption profile of 3C\,58 is a part of that 
bubble moving
towards us. The emission profiles show bright emission peaks for all major absorption
components indicating that those are major parts of the ISM. The only exception
is the absorption feature at $-41$\,km~s$^{-1}$, which has no obvious 
related emission component. There is some enhanced H~{\sc{i}} emission
around $-41$\,km~s$^{-1}$ in off-positions 2 and 3 (see Fig.~\ref{abs}),
but only the absorption profile in position 3 shows absorption. This
indicates that this absorption feature must be caused by a small 
localized cloud, which
is not part of the grand kinematic scheme of the Galaxy.

This would, however, be typical for the broken cap of an expanding bubble, 
which are very difficult to
detect due to velocity dispersion \citep{cazz05}. However, such a cap could 
reveal
itself in absorption in particular with a bright background source like 3C\,58.
\citet{koth02b} show many examples of expanding H~{\sc{i}} shells around
compact H~{\sc{ii}} regions which are revealed by H~{\sc{i}} absorption of the
H~{\sc{ii}} region's radio continuum emission by the approaching shell.

Based on these results I will adopt a systemic velocity of about 
$-36$\,km~s$^{-1}$ for 3C\,58, which is consistent with earlier estimates.
This is independent of whether 3C\,58 is located inside the bubble or
not, because the absorption feature at $-41$\,km~s$^{-1}$ represents the
absorption by a small cloud that would not be taken into account for 
the determination of a systemic velocity. And the last absorption
feature, related to a major structure of the ISM, seen in bright 
supernova remnants is typically very close to the actual
systemic velocity of these objects \citep[e.g.][]{koth03}.

\section{The Distance to 3C\,58}

\begin{figure}[!ht]
\begin{center}
 \includegraphics[bb = 55 45 535 500,width=0.48\textwidth,clip]{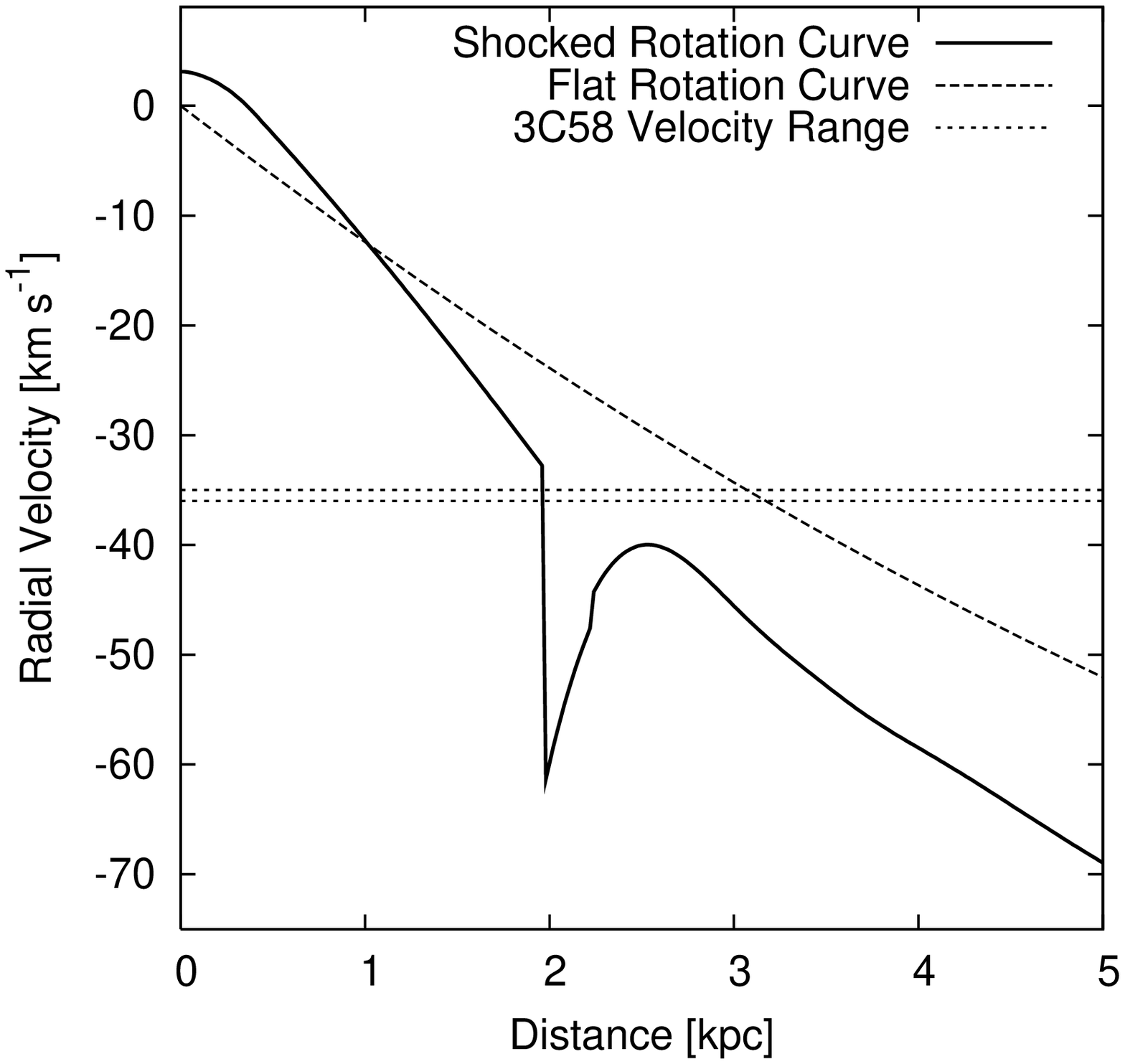}
\end{center}
\caption{Plot of the distance-velocity relation in the direction of 3C\,58
determined with the method by \citet{fost06} (solid line). The flat rotation
curve with $R_\odot = 8.5$~kpc and $v_\odot = 220$\,km~s$^{-1}$ used by
\citet{robe93} is indicated
by the dashed line. The dotted lines mark the velocity range observed for
the bubble around 3C\,58.}
\label{dvplot}
\end{figure}

The method of \citet{fost06} was used to determine a new distance-velocity relation in
the direction of 3C\,58 (Fig.~\ref{dvplot}). This is a kinematic-based distance 
tool, which accounts for not only circular motion of the Galaxy as in a
flat rotation curve model, but also for radial motions produced by spiral shocks.
To determine the distance to 3C\,58, \ion{H}{i} profiles were obtained from a
$10\degr \times 10\degr$ CGPS mosaic. This mosaic was first processed by a new
cloud filtering technique described in \citet{fost10}, which produces a 
datacube of the smooth intercloud medium by removing unwanted small clouds that may not follow
large-scale Galactic dynamics.

To produce the distance-velocity relation displayed in Fig.~\ref{dvplot}, \ion{H}{i} 
profiles along ten separate lines of sight (LOS) around 3C\,58 and a location in the 
midplane at the same longitude were fitted independently.
For the systemic velocity of 3C\,58, between $-35$ and $-36$\,km\,s$^{-1}$, the only possible location 
is at the velocity discontinuity produced by the Perseus arm spiral shock at 2\,kpc.
The distance to the spiral shock varies only slightly between the different 
lines of sight, resulting in a very small uncertainty. This leads to a distance of 
$2.0\pm 0.3$~kpc for 3C\,58.

There is a second independent method to determine a distance to 3C\,58:
relate the PWN to a nearby object or nearby objects which themselves have an independent or a more 
reliable distance estimate. The W\,3/4/5 H~{\sc{ii}} region
complex and the related SNR HB\,3 are just a few degrees away from 3C\,58. Both, HB\,3
and W\,3/4/5, have very similar systemic velocities \citep[e.g.][]{rout91}, indicating that they are all 
located in Perseus spiral arm. The distances to HB\,3, W\,5, and the nearby \ion{H}{ii} region Sh\,2-190 
were also determined by 
\citet{fost06} with the same method used here for 3C\,58. For all three objects the distance was found to
be $2.3\pm 0.5$~kpc. This is consistent with our current result. The descrepancy in
the results and errors is likely due to the application of the cloud filter in our study, which 
significantly reduces the uncertainties caused by small-scale clouds.

For W\,3 the distance was independently determined with trigonometric 
parallax to related masers to be $1.95\pm0.04$\,kpc and $2.04\pm0.07$\,kpc 
by \citet{xu06} and \citet{hach06}, respectively. This very nicely confirms the 2\,kpc
I estimated for 3C\,58 from the systemic velocity of its host stellar wind bubble.


I believe there is compelling evidence for a location of 3C\,58 in the Perseus arm 
spiral shock and from now on I will assume that 3C\,58 is located at a
distance $d = 2\pm 0.3$~kpc.

\subsection{The Progenitor of 3C\,58}

Interstellar material is compressed in the spiral shock, forms molecules, and later
stars. The shock provides the perfect environment for the formation of massive stars that later 
explode as a supernova like the progenitor star of 3C\,58. It takes a long time 
for those stars -- some $10^8$ years -- to migrate 
to a position beyond the Perseus arm spiral shock
\citep[e.g.][for similar calculations]{arvi09,koth12}. At a distance of 2\,kpc the average radius
of the bubble would be about 35\,pc. With an average expansion velocity of 
6.5\,km\,s$^{-1}$ this results in a dynamic age of $3.2\times 10^6$\,yr
for a pure stellar wind bubble produced by a single star \citep{weav77}.
Considering the age of this bubble the location of 3C\,58, about $30\arcmin$ away
from the geometric centre, requires a velocity of about 5 to 6\,km\,s$^{-1}$ perpendicular
to the line of sight for the progenitor star, which is certainly not unusual.

From the age of the stellar wind bubble we can determine approximately the type of the progenitor
star assuming that it alone was responsible for the bubble's formation. A lifetime of
$3.2\times 10^6$\,yr would point to a late O-type star with an initial mass between 
20 and 30~M$_\odot$.

\section{The Age of 3C\,58}

There is a large number of arguments against and only very few for a connection 
between the PWN 3C\,58 and the historical supernova explosion from A.D.~1181. A quite 
comprehensive list can be found in \citet[Table~3]{fese08}. The discussion
in the literature of the age of 3C\,58 has been based on an assumed distance of
3.2~kpc. I will discuss 
and re-evaluate all of the arguments, pro and contra a young age based on the 
newly determined distance of 2~kpc. The results of the discussion below
are summarized in Table~\ref{tab:age}. In the end I will try to create 
a compelling picture that explains the results of all observations and 
theoretical models.

\subsection{The Guest Star from A.D. 1181}

The strongest arguments for the connection between 3C\,58 and the Guest Star
of A.D.~1181 remain those by \citet{step71} and \citet{step02}.
The length of the visibility of the Guest Star and the similarity of the 
description in the four independent Chinese and Japanese sources 
indicate that this Guest star was the result of a supernova explosion. 
The lack of any 
other supernova remnant candidate for the event other than 3C\,58 strongly 
supports the identification of 3C\,58 with the supernova
remnant, in particular, since such a recent event should leave an
easily detectable remnant behind. The connection of 3C\,58 with the Guest
Star of A.D. 1181 implies an age of about 830~yr.

\subsubsection{Supernova Peak Brightness}

According to \citet{step02} the supernova had at its peak a visual apparent brightness
of about 0 magnitudes, maybe somewhat brighter. This would translate to an absolute brightness 
between -13.4 and -14.3~mag, using the range of possible values for visual extinction
$A_v$ \citep[1.9 to 2.8,]{shib08} and the newly determined distance of 2~kpc. This is a 
very low brightness
even if we assume an additional uncertainty of $\pm 1$~mag. 

However, other low luminosity 
supernova explosions have been observed before. 
\citet{spir09} have reviewed the literature about underluminous type II-P supernovae
and reach the conclusion that those are a quite homogenous group of supernovae with
low peak luminosities, red colours, slow moving ejecta, and small ejected $^{56}$Ni mass.
However, the presence of the stellar wind bubble around 3C\,58 indicates that the progenitor
star produced a strong stellar wind implying a supernova of type Ib or Ic, or at least 
a type IIL or IIb supernova since the progenitor must have lost a significant amount of material
through that wind, unless the bubble is unrelated to 3C\,58, in which case it could
have been a member of the underluminous type II-P group.

There have been observations of supernovae of type IIL and IIb which showed a low
visual peak brightness. A good example is SN1987a, which was a faint type IIb supernova 
explosion with an absolute visual brightness of -15.5\,mag at its peak 
\citep[][and references therein]{arne89}. The mass of the progenitor
star of SN1987a would be in approximate agreement with the progenitor mass proposed for
3C\,58 through the age of the stellar wind bubble \citep[][and references therein]{arne89}. 
However, even SN1987a was still more than 1~mag
brighter at its peak than the predicted maximum brightness for 3C\,58. And to the best 
of my knowledge there has not been a core-collapse supernova, other than a type IIP, 
observed with 
a visual peak brightness comparable to the value predicted for 3C\,58. The low absolute
visual brightness at peak is still problematic. 
It is possible that one of the main reasons
for this low brightness
is a low explosion energy, and that may explain why we still do not see the radio remnant
of the supernova explosion. Other possible reasons are small ejected $^{56}$Ni masses
\citep{spir09} and a small radius of the progenitor star, which would lead to a longer
period with large diffusion time scales. During this period the expanding star will
lose heat by adiabatic expansion. The latter reason was proposed for the low
luminosity of SN1987a \citep{arne89}.

\subsection{Spin Down Age, Neutron Star Cooling, and Other 
Pulsar-Related Arguments}

\subsubsection{Spin Down Age}
\citet{murr02} found a pulsar spin down age, also called the characteristic
age, of $\tau = 5380$~yr for J\,0205+6449, the pulsar inside 3C\,58
($\tau = \frac{P}{2\,\dot{P}}$, $P$: pulsar period). 
However, a large discrepancy between the characteristic age and the real age
is not unusual for young pulsars, and \citet{murr02} point out that their result
is not in disagreement with the connection to the supernova of
A.D. 1181. The characteristic age is only a good estimate of the
real age if the intrinsic period $P_0$ is significantly shorter than the current
period. There are many other examples of pulsars
with a characteristic age significantly larger than their actual age. One 
extreme example
would be the SNR G11.2$-$0.3 which is linked to a historically observed supernova
explosion from 386~A.D. \citep{step02} and hosts the pulsar J1811$-$1925. This
pulsar is a little over 1600~yr old and has a characteristic age of 23,300 yr 
\citep{tori99}.
The radio pulsar wind nebula in 
G11.2$-$0.3 is
much fainter than 3C\,58 \citep{koth01}, which probably can be explained by 
the much
lower energy loss rate resulting from a weaker magnetic field at birth. 

\subsubsection{Neutron Star Cooling}
\citet{slan02} fitted neutron star cooling models to CHANDRA X-ray 
observations of J\,0205+6449 and calculated strong upper limits for the 
thermal emission originating from the surface of the neutron star in 
3C\,58. These fall far below predictions for standard neutron star 
cooling. However, this does not necessarily imply a larger age for the pulsar in 
3C\,58 but it implies the presence of exotic cooling processes no 
matter whether we assume an age of 830~yr or any other age proposed for 
3C\,58 in the literature. The minimum age for 3C\,58 if cooling were to follow
standard models is about 
25,000~yr \citep[see Fig.~4 in][]{slan02}, far beyond any age that has
been proposed for 3C\,58. Exotic cooling processes must be in operation
in this pulsar, and this result does not favour any of the proposed ages.

\subsubsection{Glitches}

\citet{livi09} found two large spin up glitches in a long-term timing analysis 
of the pulsar in 3C\,58. They point out that such glitches are more
typical for pulsars with characteristic ages between 5 and 10~kyr, but suggest that
glitches could be related to the surface temperature of the neutron star
rather than the actual age, and we have just noted (Section 5.2.2) that
the pulsar in 3C\,58 has a very unusually low surface temperature.

\subsubsection{Pulsar Proper Motion}

\citet{gott07} found a circular thermal X-ray shell surrounding the pulsar 
inside 3C\,58 from {\it XMM-Newton} observations. However, the pulsar is offset
by $27''\pm 5''$ from the centroid of this shell. If 3C\,58 is the result of the 1181
supernova explosion the pulsar requires a velocity of 500~km\,s$^{-1}$ (300~km\,s$^{-1}$)
perpendicular to the line of sight to reach its current position assuming that it was
born at the very centre of this spherical shell and that its distance
is 3.2~kpc (2.0~kpc). \citet{gott07} argue that if the pulsar in 3C\,58 had the same 
transverse velocity as the Crab pulsar ($\sim 70$~km\,s$^{-1}$) it must be much older
than the event from A.D. 1181.

\citet{hobb05} have made a statistical analysis of pulsar proper motions; the
proper motion of pulsar J\,0205+6449 inside 3C\,58 is close to the average 
of their sample.
The average for normal pulsars is about 246~km\,s$^{-1}$, for pulsars in SNRs
227~km\,s$^{-1}$, and for pulsars with a characteristic age below $3\times 10^{6}$~yr
310~km\,s$^{-1}$. Hence, the pulsars's predicted proper motion at a distance
of 2~kpc supports its
connection with the supernova of A.D. 1181.
 
\subsubsection{Other Pulsar Characteristics}

\citet{shea08} say that the low optical efficiency of the emission and the
low radio to X-ray luminosity ratio observed
from the pulsar inside 3C\,58
indicate a low age, when compared with other pulsars
detected at optical wavelength. \citet{abdo09} point out that the ratio 
between X-ray and $\gamma$-ray luminosity of that pulsar is more similar 
to the middle 
aged Vela pulsar than the young Crab pulsar and B1509-58. \citet{kuip10} say 
that the spectral properties of PSR J0205+6449 inside 3C\,58
confirm that we are dealing with a young pulsar, and suggest a real 
age between those of Vela and PSR B1509-58, favouring its 
characteristic age of 5.4 kyr over the connection to SN 1181. However,
these emission characteristics cannot be taken as a means to determine
the age of the pulsar, quite simply because we have no information about 
the pulsar's emission characteristics from the time it was born.

\subsection{PWN Evolution and Energetics}

\citet{chev04} discussed evolutionary models of the expansion of the 3C\,58 
pulsar wind
nebula into the expanding ejecta of a supernova explosion. He assumed a 
power law density distribution for the ejecta into which the PWN is 
expanding and that the ejecta is swept up into a thin spherically symmetric 
shell. He finds for the ejecta mass:

\begin{equation}
\label{ejecta}
M_{ej}~[{\rm M}_\odot] = 3.5 (7.0)\,\dot{E}_{38}^{0.508}\,E_{51}^{0.492}\,R_{p}^{-2}\,t_3^{2.508}
= 2.0(4.0)\,R_{p}^{-2}\,t_3^{2.508}
\end{equation}

and for the swept-up mass:

\begin{equation}
\label{sweptup}
M_{sw}~[{\rm M}_\odot] = 0.17(0.34)\,\dot{E}_{38}\,R_{p}^{-2}\,t_3^3
= 0.054(0.108)\,R_{p}^{-2}\,t_3^3.
\end{equation}

Here, $\dot{E}_{38}$ is the energy loss rate of the pulsar in 
$10^{38}$~erg\,s$^{-1}$, $E_{51}$ the explosion energy in $10^{51}$~erg, $R_{p}$ 
the PWN radius in pc, and $t_3$ the age of the PWN in $10^3$~yr. 

For
the explosion energy I assume the canonical value of $10^{51}$~erg and
for $\dot{E}_{38}$ the average energy loss rate of the pulsar since birth
of $3.2\,10^{37}$~erg\,s$^{-1}$, assuming an age of 830~yr. The lower and
upper limits (in brackets) represent maximum and minimum coupling between 
the pulsar wind bubble and the swept up material, respectively \citep{chev04}.
In \citet{chev04} the theoretical results for swept up and ejecta mass and 
internal energy are compared with those from observations. All of these results rely 
heavily on the assumed distance. 

In the following subsections I apply these equations to various aspects of the 3C\,58 
situation, and I emphasize the major changes in the interpretation that flow from adopting
the distance 2~kpc as opposed to 3.2~kpc.

\subsubsection{Ejecta Mass}

With equation~\ref{ejecta}, a distance of 3.2 kpc, and an age of 830~yr the ejecta mass 
results in 0.1 - 0.2~M$_\odot$ for 3C\,58. This changes
to $M_{ej} = $0.3 - 0.6~M$_\odot$ for a distance of 2~kpc. This is still a 
very low value for a core collapse supernova explosion \citep{chev04}. 
If the explosion energy is much lower
than the canonical value of $10^{51}$\,erg, which is a possible
explanation for the low peak brightness of the supernova explosion (see section
5.1.1), the ejecta mass would be again much lower.

With deep optical observations \citet{fese08} revealed two distinct kinematic 
populations of expanding optical knots within the boundary of the pulsar wind 
nebula, the slower population is interpreted as 
circumstellar mass-loss material from the 3C\,58 progenitor and the higher 
velocity population as expanding supernova ejecta forming a thick shell. This 
indicates that the density distribution the PWN
is expanding into is not well represented by a power law and the ejecta is not
entirely swept up into a thin shell. In addition \citet{fese08} reported that the 
distribution of the ejecta in their observations is not spherically symmetric but
exhibits a strong bipolar expansion pattern. Therefore the theoretical estimate 
of the ejecta mass becomes very uncertain. 

\subsubsection{Swept Up Mass}

\citet{bocc01} determined from their X-ray observations a swept-up mass
for the expanding PWN of $M_{sw} = 0.1 d_{3.2}^{2.5}$\,M$_\odot$, assuming a radius $R$ of 
$2.5\arcmin$, which results in 0.1\,M$_\odot$ for a distance $d$ of 3.2\,kpc. 

We can calculate the ratio between the mass of  observed and theoretical 
(Equation~\ref{sweptup}) swept up material to be: $18(9)\,d_{3.2}^5\,t_{830}^{-3}$.
Hence, for an age of 830~yr and the distance of 3.2~kpc the observed swept up mass
is about 10-20 times higher than the theoretical one. Even with the 
relatively high uncertainty of this kind of calculation the large discrepancy 
suggests a much larger age for 3C\,58. For a distance of 2\,kpc this ratio decreases
to 0.9 - 1.7 so that the two results for the swept-up mass now agree within the
uncertainties.
According to \citet{chev04} the theoretical estimate of the swept up mass
does not rely strongly on the distribution of the ejecta. This is easily
understandable since in principle the equation calculates the mass that could have been
swept up based on the available energy. To extrapolate the total amount of
ejecta from this of course relies heavily on the assumed distribution of
the ejecta material and makes the theoretical estimate of the ejecta mass
very uncertain.

The strong dependence of the theoretically calculated swept-up mass on the 
age now actually pretty much negates an age significantly larger than 
830 yr. For an age of 1500~yr the ratio between observed and theoretically calculated
swept up mass decreases to 0.1 - 0.2, which despite the uncertainties in theoretical
estimates of that kind makes an age significantly higher than 830~yr very unlikely.

\subsubsection{Energy}

Another argument of \citet{chev04} against an age of 830\,yr is that the 
minimum energy $E_{min}$ required to produce the observed synchrotron emission is
higher than the total energy released by the pulsar. 
$E_{min}$ is about $10^{48}$\,erg assuming equipartition 
between the energy in the magnetic field and the relativistic particles 
with $E_{min} \sim d^\frac{17}{7}$ \citep{chev04}. The total energy $E_{tot}$ released 
by the pulsar into the nebula, however, can be approximated by 
$E_{tot} \approx \dot{E} t = 7\times 10^{47}$ \citep{chev04}
assuming $t = 830$\,yr. $E_{tot}$ is distance independent. We can calculate 
this value a little more accurately by integrating the equation for
the evolution of the energy loss rate, $\dot{E}$, of a
pulsar over its lifetime. $\dot{E}$ is defined by \citep[e.g.][]{paci73}:  
\begin{equation}
\label{edot}
\dot{E}(t) = \frac{\dot{E_0}}{(1+\frac{t}{\tau_0})^\beta}. 
\end{equation} 
Here
$\beta$ is related to the nature of the braking torque ($\beta =
\frac{n + 1}{n-1}$, and $n$ is the braking index of the pulsar), 
$t$ is the real
age of the pulsar, $\dot{E_0}$ is its energy loss rate
at the time it was born, and $\tau_0$ is its intrinsic characteristic age. 
Integrating 
equation~\ref{edot} over the lifetime
of 3C\,58 assuming an age of
830~\,yr and a dipolar magnetic field results in 
$E_{tot}=8.35\times 10^{47}$\,erg. This
is still significantly lower than $E_{min}$. However, for a distance of 
2\,kpc $E_{min}$
decreases to $3.2\times 10^{47}$\,erg, which is now much lower than the total 
energy released by
the pulsar, which remains unchanged. 

\subsubsection{MHD simulations}

\citet{swal03} performed MHD simulations of a magnetized pulsar wind
interacting with either a uniform interstellar medium or freely expanding
ejecta. For 3C\,58 he found that the observed radio expansion 
velocity does not agree with the radius and the age
implied by the connection with the supernova of A.D.~1181. \citet{swal03} 
suggest that there are two ways out of this. One would be a much larger age. The
second possibility is that 3C\,58 is being compressed by a reverse shock. 
According to \citet{swal03} this
would not only explain the decrease in expansion velocity, but also the radio 
flux density increase of 3C\,58 \citep{gree87}, a very low cooling break, and it would decrease
the wind termination shock. The ratio between the wind termination shock
and the PWN radius is remarkably small in 3C\,58 \citep{frai93}. 
The question remains, why do we not detect a radio bright shell 
corresponding to the blast-wave of the supernova if indeed the PWN is compressed 
by a reverse shock? And why don't we detect thermal X-ray emission from
the entire ejecta mass?

\subsection{Expansion Studies}

\begin{figure}[ht]
\begin{center}
 \includegraphics[bb = 55 45 530 480,width=0.48\textwidth,clip]{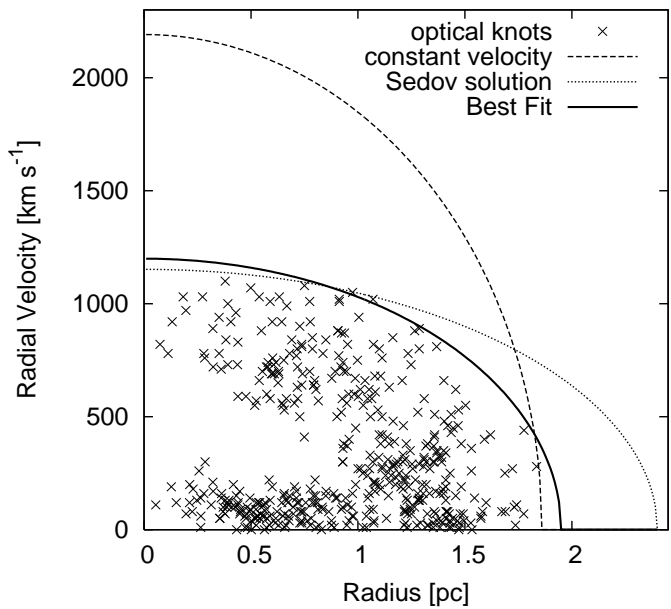}
\end{center}
\caption{Plot of the optical knots found by \citet{fese08} in a
radius-velocity diagram. The radius is the distance of the knots from
the central pulsar projected to the plane of the sky, assuming a distance 
of 2~kpc for 3C\,58. The observed absolute radial velocity is shown on the 
y-axis. The dashed line represents the outer edge of this distribution assuming
no deceleration since birth and a connection to SN\,1181. The dotted line represents
the outer edge of this distribution for Sedov expansion \citep{sedo59} and a 
connection to SN\,1181. The solid line represents
the best fit outer edge of this distribution.}
\label{exp}
\end{figure}

A supernova explosion is a one-time event. 
For any expansion study of a supernova remnant observed features related to this event
could have been decelerated since the time of the explosion but not 
accelerated. 
This may not be true for synchrotron emitting 
filaments related to a pulsar wind nebula, because those have a 
continuous source of energy in the central pulsar. 
Hence, the interpretation of expansion 
studies of synchrotron filaments inside a pulsar wind nebula
is not straightforward.

A comparison of the two major expansion studies in optical 
\citep[$\Rightarrow$ age $t \approx 3000$\,yr]{fese08} and radio 
\citep[$\Rightarrow t \approx 7000$\,yr]{biet06} already implies 
significant deceleration for the synchrotron filaments relative 
to the optically observed knots unless the energy released by the 
pulsar significantly accelerated all the supernova ejecta. In this
context -- given the age approximations of the two expansion studies -- 
significantly means that more than 80\,\% of the kinetic energy 
of the optical knots must have been provided by the pulsar.
A comparison of the typical explosion energy of $10^{51}$\,erg 
released in a supernova with the approximate energy released by the 
pulsar since birth of $8.35\times 10^{47}$\,erg (see above) negates 
that possibility entirely. Hence, we can assume that the low expansion
velocity observed for the radio synchrotron filaments is the result
of significant deceleration.

Most of the optical filaments are created by material accelerated 
through the supernova explosion. Because those filaments could have 
been decelerated but not accelerated' a simple averaging of the 
expansion velocity of all the filaments would not necessarily lead 
to a good age estimate unless the scatter is entirely produced 
by uncertainties in the observations or systematic errors, which is 
certainly not the case in the careful study of \citet{fese08}. The fastest 
filament should be taken, since it presumably shows the lowest 
deceleration. Or better in this case would be the ``youngest'' filament 
which is the filament that indicates the lowest age by simply dividing
its distance from the pulsar by its velocity. There is quite 
a wide spread in velocity in the optical study by \citet{fese08}, even
among those knots that are, according to that study, related to the
explosion and not to the circumstellar material ejected by the progenitor star's
wind. This again
indicates that a lot of the emitting material must have been 
decelerated. Therefore the assumption of insignificant deceleration for the
optical knots is invalid as well.
This weakens the case for a large age based on the optical and 
radio expansion studies. 

In Fig.~\ref{exp} I display a diagram of the optical knots constructed from the
data by
\citet{fese08}. The absolute radial velocity is plotted as a function of radius
from the central pulsar. This radius was calculated assuming that 3C\,58 is 2~kpc 
away from the sun. Since the optical knots can have been decelerated but
not accelerated any valid expansion law for the PWN has to encompass all
of these knots.

In Fig.~\ref{exp} I also display several possible expansion laws. For
all a spherical geometry has been adopted. Assuming
a connection to the supernova from 1181~A.D. the resulting best fitting functions
to encompass all expanding knots do not seem to fit very well. Free expansion
leads to a velocity-radius ratio which is too high and for the Sedov solution this
ratio is too low. The best fit expansion law indicated by the solid line represents
either free expansion for an age of about 1600~yr or Sedov expansion for an age of 
about 630~yr. Sedov expansion would result in about 0.012 for the ratio of explosion energy
$E_0$ ($10^{51}$~erg) and the ambient density $n_0$ (cm$^{-3}$), which implies either 
a very high ambient density of a very low explosion energy. 

Free expansion gives only an upper limit for the age in any case since we neglect any kind
of deceleration and the Sedov age can only be a lower limit, since the ejecta mass is
neglected. Therefore the age of 3C\,58 must be between 630 and 1600~yr. We must be in the
transition between free and Sedov expansion where neither law is sufficiently accurate to
describe the expansion.
The large spread in radial velocities for the optical knots indicates a lower rather than
a larger age, since it indicates that a lot of the ejecta must have been 
decelerated already. But in principle the 
results support any age between 630 and 1600~yr. More accurate calculations are
unfortunately not possible since we lack information about the total mass 
of the supernova ejecta and the explosion energy.

\subsection{Radio Flux Evolution}

A very strong argument for a young age is the fact that the radio flux density 
of 3C\,58 at 408~MHz increased between 1967 and 1986 at a rate of 
$0.32\pm0.13$~\%~yr$^{-1}$ \citep{gree87}. 
This is not an insignificant amount considering that the pulsar has been pumping
energy into the nebula since its birth and at a decreasing rate. The radio 
synchrotron emission of 3C\,58 depends on the energy
released by the pulsar into the nebula. Increasing radio flux implies that
the energy in magnetic field and relativistic particles  
is still being generated by the pulsar faster than it is being radiated away
by the nebula.


Although it is not easy to directly link the radio synchrotron emission to the 
total amount of energy in the nebula by
a simple equation, we can find an estimate for the age by
assuming equipartition between relativistic particles and magnetic field. In this
case we find that the flux density $S$ depends on the minimum energy $E_{min}$:
$S \sim E_{min}^\frac{7}{4}$. 

\begin{figure}[ht]
\begin{center}
 \includegraphics[bb = 55 50 540 485,width=0.48\textwidth,clip]{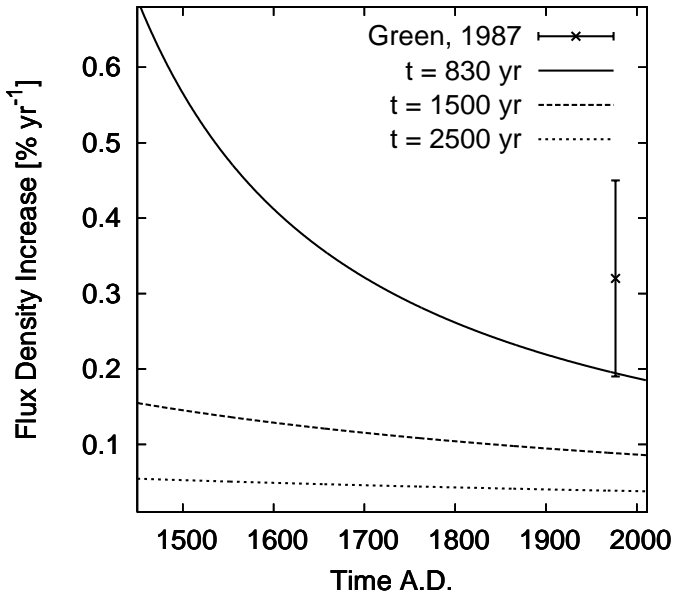}
\end{center}
\caption{Plot of the rate of radio flux density increase of 3C\,58 
over the last few hundred years predicted for different ages for the PWN. 
Negligible energy loss over its lifetime has been assumed. The flux 
density increase observed by \citet{gree87} is indicated.}
\label{eev}
\end{figure}

A flux density increase of $0.32 \pm 0.13$~\%~yr$^{-1}$ implies an
energy increase between about 0.11 and 0.26~\%~yr$^{-1}$. 
With equation~\ref{edot} this results in an age range between 350 and 
810~yr \citep[for the time range discussed in][]{gree87}, 
assuming no energy loss and a dipolar magnetic field for the
pulsar. Constant or
increasing energy loss would decrease that age range. In Fig.~\ref{eev} I
displayed the rate of radio flux density increase of 3C\,58 
over the last few hundred years predicted for different possible ages
for 3C\,58 assuming negligible energy loss. Only a current age of 830~yr would
be located within the error bars observed by \citet{gree87}.
However, even for 830~yr the presence of an additional energy source is more 
likely. This additional energy source could be the reverse shock produced by 
the interaction of the supernova shock-wave with circumstellar material,
which would be required according to the MHD simulations of \citet{swal03}
(see also Section 5.3.4) to explain the observations. For any
larger age the presence of a significant reverse shock would be essential. This
would imply the remnant has already entered Sedov phase, which implies significant
deceleration, which would indicate that the ages derived from expansion studies
are significant overestimates. A connection to the supernova of 1181~A.D. is
the only possibility to make all the observations consistent.

\begin{table}
\caption{\label{tab:age}Arguments Pro and Contra 830~yr. $\surd$ indicates that this
age is supported and $\times$ that it is not supported by the results of that method.}
\centerline{
\begin{tabular}{lcccc}
\hline \hline
Method & 830~yr & 1500~yr & $\ge 2500$~yr & Section  \\ \hline
A.D. 1181 Gueststar & $\surd$ & $\times$ & $\times$ & 5.1  \\
Peak Brigthness of SN\,1181 & $\times$ & $\surd$ & $\surd$ & 5.1.1  \\
$\tau = 5380$~yr & $\surd$ & $\surd$ & $\surd$ & 5.2.1  \\
NS Cooling & $\surd$ & $\surd$ & $\surd$ & 5.2.2  \\
Glitches & $\surd$ & $\surd$ & $\surd$ & 5.2.1  \\
Proper Motion & $\surd$ & $\surd$ & $\surd$ & 5.2.3  \\
Ejecta Mass & $\surd$ & $\surd$ &  $\surd$ &5.3.1  \\
Swept Up Mass & $\surd$ & $\times$ & $\times$ & 5.3.2  \\
Energetics & $\surd$ & $\surd$ & $\surd$ & 5.3.3  \\
MHD Simulations & $\surd$ & $\surd$ & $\surd$ & 5.3.4  \\
Radio Expansion & $\surd$ & $\surd$ & $\surd$ & 5.4  \\
Optical Expansion & $\surd$ & $\surd$ & $\times$ & 5.4  \\
Radio Flux Evolution & $\surd$ & $\times$ & $\times$ & 5.5  \\ \hline
\end{tabular}}
\end{table}

\section{The Story 3C\,58, a Proposal}

The progenitor star of 3C\,58 must have been a massive star, which was of late O-type during
its Main Sequence phase. Its strong stellar wind carved out the large stellar wind bubble
we find in H~{\sc{i}}. The dynamic age of that bubble of about $3.2\times 10^6$~yr
indicates a mass between 20 and 30~M$_\odot$ for the progenitor star when it was formed,
assuming the bubble was blown by that single star.
In its late evolutionary phase this star must have been a red giant, supergiant star or
even a blue super giant similar to the progenitor of SN1987a,
as indicated by the production of a massive low velocity
wind which is implied by the slowly expanding optical knots found by \citet{fese08}.
When the star exploded it left the neutron star that powers 3C\,58 behind and the 
shock-wave of the explosion expanded into the very low density interior of the stellar
wind bubble which was partly filled with high density clouds expelled by the progenitor
star in its wind before it exploded. Consequently, the early expansion of the supernova shock wave 
has
to be treated as expanding into a cloudy instead of a homogeneous medium. This explains the 
not limb-brightened partly centrally peaked thermal X-ray emission which was observed by 
\citet[their Fig.~8, right panel]{gott07} and the lack of non-thermal radio emission 
related to the supernova remnant (not the PWN) \citep[e.g.][]{whit91}. The low brightness
of the supernova explosion at the peak of its light curve can be explained by a low
explosion energy, small mass of ejected $^{56}$Ni, a small progenitor radius, or various
other situations.  The interaction
with the circumstellar material is sapping the shock wave of its energy and generates
reverse shocks moving back into the interior. These are decelerating the ejecta, creating 
the large velocity spread inside a thick shell. This reverse shock is also decelerating and 
compressing the PWN thereby increasing its energy. This
explains the continuous significant increase of radio flux from the PWN.

Although the new distance estimate explains many of the unusual characteristics found for
3C\,58 a few problems still remain:
\begin{itemize}
\item Why did the supernova explosion that created 3C\,58 have such a low peak brightness?
\item Where is the radio remnant of the supernova explosion?
\item Why is the extrapolated mass of the ejecta so low?
\item Why does the neutron star display such a low surface temperature?
\end{itemize}

\section{Summary}

I have derived a new more reliable distance of 2\,kpc to the PWN 3C\,58, by means of 
H~{\sc{i}} absorption in combination with a newly determined distance-velocity relation 
and by relating this PWN to a nearby H~{\sc{ii}} region and SNR complex of well known 
distance. This new distance changes many characteristics of this PWN quite
dramatically. The new results point once again to the historical
connection of 3C\,58, because only an age of 830~yr makes most of the observations
and theoretical calculations consistent. 

\begin{acknowledgements}
I would like to thank Tyler Foster for providing me with a distance-velocity
diagram in the direction of 3C\,58 using his modeling technique of the Galactic hydrogen 
distribution. I would also like to thank Tom Landecker for careful reading of this
manuscript. The Dominion Radio Astrophysical Observatory is a National 
Facility
operated by the National Research Council of Canada. The Canadian Galactic
Plane Survey is a Canadian project with international partners, and is
supported by the Natural Sciences and Engineering Research Council
(NSERC).
\end{acknowledgements}

\bibliographystyle{aa}
\bibliography{kothes}

\begin{thebibliography}{48}
\expandafter\ifx\csname natexlab\endcsname\relax\def\natexlab#1{#1}\fi

\bibitem[{{Abdo} {et~al.}(2009){Abdo}, {Ackermann}, {Ajello}, {Atwood},
  {Axelsson}, {Baldini}, {Ballet}, {Barbiellini}, {Bastieri}, {Baughman},
  {Bechtol}, {Bellazzini}, {Berenji}, {Blandford}, {Bloom}, {Bonamente},
  {Borgland}, {Bouvier}, {Bregeon}, {Brez}, {Brigida}, {Bruel}, {Burnett},
  {Caliandro}, {Cameron}, {Camilo}, {Caraveo}, {Casandjian}, {Cecchi}, {{\c
  C}elik}, {Chekhtman}, {Cheung}, {Chiang}, {Ciprini}, {Claus}, {Cognard},
  {Cohen-Tanugi}, {Conrad}, {Dermer}, {de Angelis}, {de Palma}, {Digel},
  {Dormody}, {do Couto e Silva}, {Drell}, {Dubois}, {Dumora}, {Edmonds},
  {Espinoza}, {Farnier}, {Favuzzi}, {Focke}, {Frailis}, {Freire}, {Fukazawa},
  {Fusco}, {Gargano}, {Gehrels}, {Germani}, {Giebels}, {Giglietto}, {Giordano},
  {Glanzman}, {Godfrey}, {Grenier}, {Grondin}, {Grove}, {Guillemot}, {Guiriec},
  {Hanabata}, {Harding}, {Hayashida}, {Hays}, {Hobbs}, {Hughes},
  {J{\'o}hannesson}, {Johnson}, {Johnson}, {Johnson}, {Johnson}, {Johnston},
  {Kamae}, {Kaspi}, {Katagiri}, {Kataoka}, {Kawai}, {Keith}, {Kerr},
  {Kn{\"o}dlseder}, {Kramer}, {Kuehn}, {Kuss}, {Lande}, {Latronico},
  {Lemoine-Goumard}, {Livingstone}, {Longo}, {Loparco}, {Lott}, {Lovellette},
  {Lubrano}, {Lyne}, {Makeev}, {Manchester}, {Marelli}, {Mazziotta}, {McEnery},
  {Meurer}, {Michelson}, {Mitthumsiri}, {Mizuno}, {Moiseev}, {Monte},
  {Monzani}, {Morselli}, {Moskalenko}, {Murgia}, {Nolan}, {Nuss}, {Ohsugi},
  {Omodei}, {Orlando}, {Ormes}, {Paneque}, {Panetta}, {Parent}, {Pelassa},
  {Pepe}, {Pesce-Rollins}, {Pierbattista}, {Piron}, {Porter}, {Rain{\`o}},
  {Rando}, {Ransom}, {Razzano}, {Reimer}, {Reimer}, {Reposeur}, {Ritz},
  {Rochester}, {Rodriguez}, {Romani}, {Ryde}, {Sadrozinski}, {Sanchez},
  {Sander}, {Saz Parkinson}, {Sgr{\`o}}, {Siskind}, {Smith}, {Smith},
  {Spandre}, {Spinelli}, {Stappers}, {Striani}, {Strickman}, {Suson}, {Tajima},
  {Takahashi}, {Tanaka}, {Thayer}, {Thayer}, {Theureau}, {Thompson},
  {Thorsett}, {Tibaldo}, {Torres}, {Tosti}, {Tramacere}, {Uchiyama}, {Usher},
  {Van Etten}, {Vasileiou}, {Vilchez}, {Vitale}, {Waite}, {Wang}, {Watters},
  {Weltevrede}, {Winer}, {Wood}, {Ylinen}, \& {Ziegler}}]{abdo09}
{Abdo}, A.~A., {Ackermann}, M., {Ajello}, M., {et~al.} 2009, \apjl, 699, L102

\bibitem[{{Arnett} {et~al.}(1989){Arnett}, {Bahcall}, {Kirshner}, \&
  {Woosley}}]{arne89}
{Arnett}, W.~D., {Bahcall}, J.~N., {Kirshner}, R.~P., \& {Woosley}, S.~E. 1989,
  \araa, 27, 629

\bibitem[{{Arvidsson} {et~al.}(2009){Arvidsson}, {Kerton}, \&
  {Foster}}]{arvi09}
{Arvidsson}, K., {Kerton}, C.~R., \& {Foster}, T. 2009, \apj, 700, 1000

\bibitem[{{Bietenholz}(2006)}]{biet06}
{Bietenholz}, M.~F. 2006, \apj, 645, 1180

\bibitem[{{Bocchino} {et~al.}(2001){Bocchino}, {Warwick}, {Marty}, {Lumb},
  {Becker}, \& {Pigot}}]{bocc01}
{Bocchino}, F., {Warwick}, R.~S., {Marty}, P., {et~al.} 2001, \aap, 369, 1078

\bibitem[{{Cazzolato} \& {Pineault}(2005)}]{cazz05}
{Cazzolato}, F. \& {Pineault}, S. 2005, \aj, 129, 2731

\bibitem[{{Chevalier}(2004)}]{chev04}
{Chevalier}, R.~A. 2004, Advances in Space Research, 33, 456

\bibitem[{{Fesen} {et~al.}(2008){Fesen}, {Rudie}, {Hurford}, \&
  {Soto}}]{fese08}
{Fesen}, R., {Rudie}, G., {Hurford}, A., \& {Soto}, A. 2008, \apjs, 174, 379

\bibitem[{{Foster} \& {Cooper}(2010)}]{fost10}
{Foster}, T. \& {Cooper}, B. 2010, in Astronomical Society of the Pacific
  Conference Series, Vol. 438, Astronomical Society of the Pacific Conference
  Series, ed. {R.~Kothes, T.~L.~Landecker, \& A.~G.~Willis}, 16--+

\bibitem[{{Foster} \& {MacWilliams}(2006)}]{fost06}
{Foster}, T. \& {MacWilliams}, J. 2006, \apj, 644, 214

\bibitem[{{Foster} {et~al.}(2004){Foster}, {Routledge}, \& {Kothes}}]{fost04}
{Foster}, T., {Routledge}, D., \& {Kothes}, R. 2004, \aap, 417, 79

\bibitem[{{Frail} \& {Moffett}(1993)}]{frai93}
{Frail}, D.~A. \& {Moffett}, D.~A. 1993, \apj, 408, 637

\bibitem[{{Gotthelf} {et~al.}(2007){Gotthelf}, {Helfand}, \&
  {Newburgh}}]{gott07}
{Gotthelf}, E.~V., {Helfand}, D.~J., \& {Newburgh}, L. 2007, \apj, 654, 267

\bibitem[{{Green}(1987)}]{gree87}
{Green}, D.~A. 1987, \mnras, 225, 11P

\bibitem[{{Green} \& {Gull}(1982)}]{gree82}
{Green}, D.~A. \& {Gull}, S.~F. 1982, \nat, 299, 606

\bibitem[{{Hachisuka} {et~al.}(2006){Hachisuka}, {Brunthaler}, {Menten},
  {Reid}, {Imai}, {Hagiwara}, {Miyoshi}, {Horiuchi}, \& {Sasao}}]{hach06}
{Hachisuka}, K., {Brunthaler}, A., {Menten}, K.~M., {et~al.} 2006, \apj, 645,
  337

\bibitem[{{Higgs} \& {Tapping}(2000)}]{higg00}
{Higgs}, L.~A. \& {Tapping}, K.~F. 2000, \aj, 120, 2471

\bibitem[{{Hobbs} {et~al.}(2005){Hobbs}, {Lorimer}, {Lyne}, \&
  {Kramer}}]{hobb05}
{Hobbs}, G., {Lorimer}, D.~R., {Lyne}, A.~G., \& {Kramer}, M. 2005, \mnras,
  360, 974

\bibitem[{{Kothes}(2010)}]{koth10}
{Kothes}, R. 2010, in Astronomical Society of the Pacific Conference Series,
  Vol. 438, Astronomical Society of the Pacific Conference Series, ed.
  {R.~Kothes, T.~L.~Landecker, \& A.~G.~Willis}, 347

\bibitem[{{Kothes} \& {Foster}(2012)}]{koth12}
{Kothes}, R. \& {Foster}, T. 2012, \apjl, 746, L4

\bibitem[{{Kothes} \& {Kerton}(2002)}]{koth02b}
{Kothes}, R. \& {Kerton}, C.~R. 2002, \aap, 390, 337

\bibitem[{{Kothes} \& {Reich}(2001)}]{koth01}
{Kothes}, R. \& {Reich}, W. 2001, \aap, 372, 627

\bibitem[{{Kothes} {et~al.}(2003){Kothes}, {Reich}, {Foster}, \&
  {Byun}}]{koth03}
{Kothes}, R., {Reich}, W., {Foster}, T., \& {Byun}, D.-Y. 2003, \apj, 588, 852

\bibitem[{{Kothes} {et~al.}(2002){Kothes}, {Uyaniker}, \& {Yar}}]{koth02}
{Kothes}, R., {Uyaniker}, B., \& {Yar}, A. 2002, \apj, 576, 169

\bibitem[{{Kuiper} {et~al.}(2010){Kuiper}, {Hermsen}, {Urama}, {den Hartog},
  {Lyne}, \& {Stappers}}]{kuip10}
{Kuiper}, L., {Hermsen}, W., {Urama}, J.~O., {et~al.} 2010, \aap, 515, A34+

\bibitem[{{Landecker} {et~al.}(2000){Landecker}, {Dewdney}, {Burgess}, {Gray},
  {Higgs}, {Hoffmann}, {Hovey}, {Karpa}, {Lacey}, {Prowse}, {Purton}, {Roger},
  {Willis}, {Wyslouzil}, {Routledge}, \& {Vaneldik}}]{land00}
{Landecker}, T.~L., {Dewdney}, P.~E., {Burgess}, T.~A., {et~al.} 2000, \aaps,
  145, 509

\bibitem[{{Livingstone} {et~al.}(2009){Livingstone}, {Ransom}, {Camilo},
  {Kaspi}, {Lyne}, {Kramer}, \& {Stairs}}]{livi09}
{Livingstone}, M.~A., {Ransom}, S.~M., {Camilo}, F., {et~al.} 2009, \apj, 706,
  1163

\bibitem[{{Murray} {et~al.}(2002){Murray}, {Slane}, {Seward}, {Ransom}, \&
  {Gaensler}}]{murr02}
{Murray}, S.~S., {Slane}, P.~O., {Seward}, F.~D., {Ransom}, S.~M., \&
  {Gaensler}, B.~M. 2002, \apj, 568, 226

\bibitem[{{Pacini} \& {Salvati}(1973)}]{paci73}
{Pacini}, F. \& {Salvati}, M. 1973, \apj, 186, 249

\bibitem[{{Roberts} {et~al.}(1993){Roberts}, {Goss}, {Kalberla}, {Herbstmeier},
  \& {Schwarz}}]{robe93}
{Roberts}, D.~A., {Goss}, W.~M., {Kalberla}, P.~M.~W., {Herbstmeier}, U., \&
  {Schwarz}, U.~J. 1993, \aap, 274, 427

\bibitem[{{Roberts}(1972)}]{robe72}
{Roberts}, Jr., W.~W. 1972, \apj, 173, 259

\bibitem[{{Routledge} {et~al.}(1991){Routledge}, {Dewdney}, {Landecker}, \&
  {Vaneldik}}]{rout91}
{Routledge}, D., {Dewdney}, P.~E., {Landecker}, T.~L., \& {Vaneldik}, J.~F.
  1991, \aap, 247, 529

\bibitem[{{Schmidt}(1965)}]{schm65}
{Schmidt}, M. 1965, in Galactic Structure, ed. {A.~Blaauw \& M.~Schmidt},
  513--+

\bibitem[{{Sedov}(1959)}]{sedo59}
{Sedov}, L.~I. 1959, {Similarity and Dimensional Methods in Mechanics} (New
  York: Academic Press)

\bibitem[{{Shearer} \& {Neustroev}(2008)}]{shea08}
{Shearer}, A. \& {Neustroev}, V.~V. 2008, \mnras, 390, 235

\bibitem[{{Shibanov} {et~al.}(2008){Shibanov}, {Lundqvist}, {Lundqvist},
  {Sollerman}, \& {Zyuzin}}]{shib08}
{Shibanov}, Y.~A., {Lundqvist}, N., {Lundqvist}, P., {Sollerman}, J., \&
  {Zyuzin}, D. 2008, \aap, 486, 273

\bibitem[{{Slane} {et~al.}(2002){Slane}, {Helfand}, \& {Murray}}]{slan02}
{Slane}, P.~O., {Helfand}, D.~J., \& {Murray}, S.~S. 2002, \apjl, 571, L45

\bibitem[{{Spiro} \& {Pastorello}(2009)}]{spir09}
{Spiro}, S. \& {Pastorello}, A. 2009, in American Institute of Physics
  Conference Series, Vol. 1111, American Institute of Physics Conference
  Series, ed. {G.~Giobbi, A.~Tornambe, G.~Raimondo, M.~Limongi,
  L.~A.~Antonelli, N.~Menci, \& E.~Brocato}, 460--463

\bibitem[{{Stephenson}(1971)}]{step71}
{Stephenson}, F.~R. 1971, \qjras, 12, 10

\bibitem[{{Stephenson} \& {Green}(2002)}]{step02}
{Stephenson}, F.~R. \& {Green}, D.~A. 2002, Historical supernovae and their
  remnants, by F.~Richard Stephenson and David A.~Green.~International series
  in astronomy and astrophysics, vol.~5.~Oxford: Clarendon Press, 2002, ISBN
  0198507666, 5

\bibitem[{{Taylor} {et~al.}(2003){Taylor}, {Gibson}, {Peracaula}, {Landecker},
  {Brunt}, {Dewdney}, {Dougherty}, {Gray}, {Higgs}, {Kerton}, {Knee}, {Kothes},
  {Purton}, {Uyan{\i}ker}, {Wallace}, {Willis}, \& {Durand}}]{tayl03}
{Taylor}, A.~R., {Gibson}, S.~J., {Peracaula}, M., {et~al.} 2003, \aj, 125,
  3145

\bibitem[{{Torii} {et~al.}(1999){Torii}, {Tsunemi}, {Dotani}, {Mitsuda},
  {Kawai}, {Kinugasa}, {Saito}, \& {Shibata}}]{tori99}
{Torii}, K., {Tsunemi}, H., {Dotani}, T., {et~al.} 1999, \apjl, 523, L69

\bibitem[{{van der Swaluw}(2003)}]{swal03}
{van der Swaluw}, E. 2003, \aap, 404, 939

\bibitem[{{Wallace} {et~al.}(1994){Wallace}, {Landecker}, \& {Taylor}}]{wall94}
{Wallace}, B.~J., {Landecker}, T.~L., \& {Taylor}, A.~R. 1994, \aap, 286, 565

\bibitem[{{Weaver} {et~al.}(1977){Weaver}, {McCray}, {Castor}, {Shapiro}, \&
  {Moore}}]{weav77}
{Weaver}, R., {McCray}, R., {Castor}, J., {Shapiro}, P., \& {Moore}, R. 1977,
  \apj, 218, 377

\bibitem[{{White} \& {Long}(1991)}]{whit91}
{White}, R.~L. \& {Long}, K.~S. 1991, \apj, 373, 543

\bibitem[{{Williams}(1973)}]{will73}
{Williams}, D.~R.~W. 1973, \aap, 28, 309

\bibitem[{{Xu} {et~al.}(2006){Xu}, {Reid}, {Zheng}, \& {Menten}}]{xu06}
{Xu}, Y., {Reid}, M.~J., {Zheng}, X.~W., \& {Menten}, K.~M. 2006, Science, 311,
  54

\end{thebibliography}

\end{document}